# Efficient diffusive mechanisms of O atoms at very low temperatures on surfaces of astrophysical interest

Emanuele Congiu,*[a] Marco Minissale,[a,c] Saoud Baouche,[a] Henda Chaabouni,[a] Audrey Moudens,[a] Stephanie Cazaux,[b] Giulio Manicò,[c] Valerio Pirronello[c] and François Dulieu[a]

At the low temperatures of interstellar dust grains, it is well established that surface chemistry proceeds via diffusive mechanisms of H atoms weakly bound (physisorbed) to the surface. Until recently, however, it was unknown whether atoms heavier than hydrogen could diffuse rapidly enough on interstellar grains to react with other accreted species. In addition, models still require simple reduction as well as oxidation reactions to occur on grains to explain the abundances of various molecules. In this paper we investigate O-atom diffusion and reactivity on a variety of astrophysically relevant surfaces (water ice of three different morphologies, silicate, and graphite) in the 6.5 – 25 K temperature range. Experimental values were used to derive a diffusion law that emphasizes that O atoms diffuse by quantum mechanical tunnelling at temperatures as low as 6.5 K. The rate of diffusion on each surface, based on modelling results, were calculated and an empirical law is given as a function of the surface temperature. Relative diffusion rates are $k_{H2Oice} > k_{sil} > k_{graph} \gg k_{expected}$. The implications of an efficient O-atom diffusion over astrophysically relevant time-scales are discussed. Our findings show that O atoms can scan any available reaction partners (e.g., either another H atom, if available, or a surface radical like O or OH) at a faster rate than that of accretion. Also, as dense clouds mature $H_2$ becomes far more abundant than H and the O/H ratio grows, the reactivity of O atoms on grains is such that O becomes one of the dominant reactive partners together with H.

## Introduction

In the cold regions of the Universe, where temperatures are lower than 10 K and densities are far weaker than those attainable on earth, a rich chemistry is initiated on the surfaces of minuscule dust particles.[1-4] The species weakly bound to the surface, most of all, play a central role in the evolution of the pristine chemistry governed by the diffusion of reactive species.[5] In space, thermal atom-addition induced chemistry occurs mostly at low temperatures (~10 K), i.e., in the innermost part of the clouds where newly formed species are protected from radiation to a great extent by dust particles. These regions are parts of collapsing envelopes that feed young stellar objects and that provide the original material from which comets and ultimately planets are made.[6] Hydrogenation of interstellar ices can induce the formation of new species in the solid phase and, therefore, it has been the topic of recent laboratory-based studies.[7-10] The efficient surface formation of the bulk of interstellar ices, i.e., water, methanol, formaldehyde, and formic acid has been demonstrated through H-atom additions of CO- and/or $O_2$-ices under interstellar relevant conditions.[11-13] In particular, the formation of the most important and abundant ice of all, amorphous solid water (ASW), proved to be the principal product of all the possible chemical pathways involving O and H atoms/molecules or the hydroxyl radical ($O_3$, $O_2$, O, OH, H and $H_2$).[14-20] So far, diffusion has only been explored experimentally for H atoms.[21,22] Nevertheless, if we assume that only H atoms are mobile and can diffuse over the dust grain surface, it becomes difficult to meet the observational evidence for $CO_2$ – the second most abundant condensed species in grain mantles – as well as for the rich molecular diversity for the general interstellar medium. $CO_2$ is believed to form in the solid phase via several energetic[23-26] and non-energetic[23-32] mechanisms, i.e., CO+OH and CO+O reactions at 10 K. The reaction CO+OH leading to $CO_2$, however, occurs in competition with $H_2O$ formation via the H+OH (barrierless) pathway,[30] which is a much faster route whenever H is the only mobile species. For this reason, the CO+OH reaction alone cannot account for the $CO_2$ ice observed in quiescent regions.[33,34] Furthermore, it seems that a sort of depth (i.e., age) segregation exists between the three most abundant ices ($H_2O$, $CO_2$, CO):[35,36] water tends to be concentrated in the layers forming the inner (and older) part of the mantles, while $CO_2$ and CO abundances increase in the outer (and more recent) layers. CO accretes onto icy grain surfaces at a rate proportional to the gas density,[37,38] almost certainly because dust grains are not cold enough in low density regions (e.g., diffuse clouds) so that CO residence time on the surface is extremely short. Conversely, why $H_2O$ (together with other fully hydrogenated species like $CH_3OH$ and $NH_3$)[35] is mostly found in the inner part (near to the silicate/carbonaceous core), and $CO_2$ in the outer layers, is not fully understood. This could actually be explained by assuming the presence of efficient diffusive processes of O atoms on the icy grains, as recent laboratory works by our group demonstrated on realistic dust grain analogues, such as water ice and silicate at very low temperatures.[39,40] Also, this scenario is consistent with the $CO_2$ formation timeline proposed in *Noble et al*,[41] that they tested through concurrent observations of CO, $CO_2$, and $H_2O$ ices.

[a] LERMA Université de Cergy-Pontoise and Observatoire de Paris, ENS, UPMC, UMR 8112 du CNRS 5, mail Gay Lussac, 95000 Cergy Pontoise cedex, France; E-mail: emanuele.congiu@u-cergy.fr
[b] Kapteyn Astronomical Institute, PO box 800, 9700AV Groningen, The Netherlands;
[c] Dipartimento di Fisica e Astronomia, Università di Catania, via Santa Sofia 64, 95123 Catania, Sicily, Italy.

Here we present new results of diffusion constants of O atoms calculated on graphite, and compare them to the diffusion constants previously calculated on amorphous silicate and three different surfaces of water ice: porous, compact amorphous, and crystalline water ice. In this comparative study, the measured diffusion constants on each surface are given as a function of surface temperature. These data are modelled to determine a simple analytical expression that accurately reproduces the diffusion constants of O atoms with surface temperature. Also, by changing the model to incorporate both a classical (Arrhenius-type) law and a quantum tunnelling description, we are able to address some key physical questions, namely what diffusive process is at play in the 6 – 25 K temperature range.

**Experimental**

The FORmation of MOLecules in the InterStellar Medium (FORMOLISM) experimental set-up has been developed with the purpose of studying the reaction and interaction of atoms and molecules on surfaces simulating dust grains under interstellar conditions (the relevance of substrate, low density, and very low temperatures ~10 K).[42] FORMOLISM is composed of an ultrahigh vacuum chamber with a base pressure of ~$10^{-10}$ mbar, a rotatable quadrupole mass spectrometer (QMS) and an oxygen-free high-conductivity copper sample holder. The sample holder is attached to the cold finger of a closed-cycle He cryostat and can be cooled to 6 K. The temperature is measured with a calibrated silicon diode clamped on the sample holder and controlled by a Lakeshore 334 controller to ±0.2 K with an accuracy of ±1 K in the 8−400 K range. The apparatus is also equipped with a reflection-absorption infrared spectroscopy (RAIRS) facility used to probe the deposited or produced species *in situ*.[7] Reactants are introduced into the vacuum chamber via two separated triply differentially pumped beam lines aimed at the cold surface. Each beam line, in its first stage, consists of an air-cooled quartz tube surrounded by a microwave cavity for dissociating select species (e.g., $H_2$, $O_2$, $N_2$). In this study, only one beam line was used to dissociate and deposit O atoms. Typical values of the dissociation rate are ~ 70%, which means that the $O/O_2$ mixture sent onto the sample is in the ratio 14/3 (i.e., 0.7·2O/0.3·$O_2$). Atoms are cooled and thermalised to 300 – 400 K upon impact with the surfaces of the quartz tube. We also determined that O atoms and $O_2$ molecules exiting the source are in their ground state $^3P$ and $X^3\Sigma_g^-$, respectively. The beam flux was calibrated by using temperature-programmed desorption (TPD) to determine the $O_2$ exposure time to saturate the $O_2$ monolayer.

Five surfaces were investigated in this study: porous ASW ($H2O_{(p)}$), non-porous ASW ($H2O_{(np)}$), crystalline ice ($H2O_{(c)}$), amorphous olivine-type silicate ($SiO_x$), and an oxidised slab of highly oriented pyrolytic graphite (HOPG). The $SiO_x$, HOPG, and other carbon-based surfaces, mimic bare dust grains in molecular clouds and have been previously used in investigations of heterogeneous chemistry on dust grain analogues. The silicate sample is amorphous in nature, as evidenced by infrared spectroscopic studies,[43] while TPD experiments reveal the surface to be non-porous on the molecular scale.[44] The HOPG surface used in the experiments is a ZYA-grade HOPG sample, which had been previuosly exposed to an O-atom beam (oxidised) to avoid surface changes during the experimental sequences. For the water substrates, ice films were grown on top of the silicate surface by spraying water vapour from a microchannel array doser located 2 cm in front of the surface. The water vapour was obtained from deionised water which had been purified by several freeze–pump–thaw cycles, carried out under vacuum. $H_2O_{(p)}$ and $H_2O_{(np)}$ mimic the ASW which comprises the bulk of interstellar ice, and $H_2O_{(c)}$ mimics the crystalline ice seen in some star-forming regions. To produce $H_2O_{(np)}$, water was dosed while the surface was held at a constant temperature of 110 K. $H_2O_{(p)}$ was grown at 10 K on top of $H_2O_{(np)}$, then the composite film was annealed to 90 K to stabilize the surface morphology before subsequent heating-cooling runs between 6.5 and 90 K. The sub-layer of H2O$_{(np)}$ has a thickness of ~ 50 ML (1 ML = $10^{15}$ molecules cm$^{-2}$) and its purpose is to isolate the ensuing H$_2$O$_{(p)}$ films from the SiO$_x$ substrate.[45] To form H$_2$O$_{(c)}$, the surface was held at 120 K during the deposition, then flash heated at 50 K min$^{-1}$ to 145 K. For each type of ice surface, the temperature was held constant until the background pressure in the chamber stabilised, before cooling it back down in the range 6 – 25 K, at which temperature O atoms were dosed onto the respective surfaces.

**Results**

In Fig. 1 we show the type of raw experimental data we obtain: O$_2$ and O$_3$ mass spectra (TPD performed at a heating rate of 10 K min$^{-1}$) are recorded after deposition of a O+O$_2$ dose on graphite at a given surface temperature (Fig. 1, left panel). O$_2$ desorbs between 27 and 50 K, while ozone desorption is observed between 55 and 75 K (directly at mass 48 a.m.u., or via the O$_2^+$ fragments at mass 32 a.m.u.). O (16 a.m.u.) desorption was never observed. The area of the peaks (proportional to the amount of the species formed on the surface) changes depending on the coverage (O+O$_2$ dose) and on the surface temperature. Fig. 1, right panel: RAIR spectrum of the $\nu_3$-ozone band at 1043.5 cm$^{-1}$ recorded at 6.5 K after deposition of 0.3 ML of O+O$_2$ on graphite at the same temperature. O$_3$ and O$_2$ yields are calculated after depositions of O+O$_2$ performed by a) varying the dose (coverage) in the range 0.05 – 1 ML and keeping the surface temperature (T$_s$) fixed, and b) dosing with a fixed amount and varying the deposition temperature between 6 and 30 K. The acquisition of two sets of data is motivated by the nature of the mechanisms through which surface reactions leading to O$_2$ and O$_3$ proceed. The two main mechanisms invoked for surface catalysis are the Langmuir–Hinshelwood (LH) mechanism, in which O$_2$ and O$_3$ form in an O-atom diffusion process, and the Eley–Rideal (ER) mechanism, or 'prompt' mechanism, in which an impinging O-atom reacts directly with an adsorbed O or O$_2$. The LH mechanism is highly dependent on the surface temperature as it affects the mobility of species on the surface. This justifies depositions of O$_2$ and O$_3$ at different surface temperatures. The ER mechanism, on the other hand, is indipendent of T$_s$, becomes more efficient with the increase in surface coverage, and therefore we also investigated the O$_2$ and O$_3$ formation at several O+$O_2$ initial doses. In two previous papers[39,40] dealing with the diffusion of oxygen atoms on water ice and amorphous silicate, we showed the complete series of TPDs performed at several O+$O_2$ coverages and at various surface temperatures. From the resulting $O_3$ and $O_2$ yields, we inferred that the $O_3/O_2$ ratio increases with initial coverage, as an incoming O-atom is more likely to find $O_2$ molecules at higher coverages, and with surface temperature because at higher temperatures the mobility of O atoms is favoured and ozone formation is more efficient.

If we focus on the diffusion dependency on $T_s$, and at coverages less than 0.5 ML – more suitable to the astrophysical context – the LH mechanism can be fairly considered the main process governing the surface reactions involving oxygen atoms and molecules. One may argue, however, that our experiments – based on linear thermal ramps starting from deposition temperature $T_s$ – can affect the diffusion of O atoms and the actual $O_2$ and $O_3$ yields. To rule this possibility out, we planned an experiment in which TPD and RAIRS results could be compared in order to confirm or discard the possible role of the heating ramp in O diffusion. In Fig. 2 we present the amount of ozone formed after depositing a given dose of O+$O_2$ on graphite. The experiment was performed using two different procedures. First, after a 0.3 ML of O+$O_2$ was dosed at various temperatures, a RAIR spectra was recorded then a TPD was started for each deposition. Blue squares in Fig. 2 indicate the TPD ozone yields after deposition of 0.3 ML of O+$O_2$ on oxidised HOPG kept at 6.5, 10, 15, 20 and 25 K. Green squares were obtained by integration of the band area at 1043.5 $cm^{-1}$ of ozone, recorded after each deposition. Second, after depositing 0.3 ML of O+$O_2$ on graphite at 6.5 K, we monitored the evolution of the ozone IR band intensity with temperature. By this second method we could estimate the formation of ozone during the thermal ramp. The result of this experiment is shown by the black circles in Fig. 2. They give the ozone IR band area as a function of temperature after a single 0.3-ML dose of O+$O_2$ at 6.5 K. For temperatures greater than 40 K, the $O_3$ band does not increase because $O_2$ desorbs and the O+$O_2$ reaction can no longer take place. The data presented in Fig. 2, in summary, show the yield of ozone formed following a given dose of O+$O_2$ deposited at different $T_s$ between 6 and 25 K ($T_s$ greater than 25 K would make $O_2$ mobile as well, hence add another degree of complexity to this study). It is clear that ozone formation efficiency grows fast with the deposition temperature (see squares in Fig. 2), while the contribution to ozone provided by the thermal ramp – which is likely to induce diffusion of the residual O atoms – is very small (see black circles in Fig. 2). In addition, if we consider the IR data at 15 K, the $O_3$ yield obtained after $O+O_2$ deposition at 15 K (green square) is much higher than the ozone yield after deposition at 6.5 K and heated to 15 K (black circle). This fact confirms that all the chemistry has occurred at the deposition temperature; and if it is true at Ts = 6.5 K, then it is the case at any other temperature higher than 6.5 K.

Experimental data are then inserted into a model composed by a series of rate equations used to simulate the $O_2$ and $O_3$ formation yields according to coverage and surface temperature. The model includes both LH and ER mechanisms, and it allows reactions to occur during the deposition phase, as well as during the heating phase (TPD). A complete account of our model is given in *Minissale et al 2014*.[40] Here we will focus on the diffusion rates k and the different methods by which they are calculated. We already alluded to the fact that reactions mostly occur during the exposure phase. The diffusion of atoms during the heating phase is small because not more than a few percent of the deposited O atoms remain available on the surface in the low coverage regime. The effect of a possible diffusion during the TPD lies within the error bars of the experimental data, and can be neglected. For this reason, in what follows, we will address only the diffusion constants at a fixed temperature for each one of the substrate investigated.

The diffusion coefficients k include two components due to quantum tunnelling and thermal motion:[46]

$$k = k_{qt} + k_{tm}$$

In our model, k can be treated as a free numerical parameter during the deposition phase at constant temperature, owing to the fact that the evolution of the coverage with time is known and provides a strong constraint. Therefore, resulting k values are a set of constants giving the diffusion rate at given temperatures, although no information can be inferred about the nature of the diffusive process. In Fig. 3 the diffusion constants k that we obtained for various substrate compositions are plotted as a function of temperature. An important finding of this comparative study is that diffusion coefficients on water ices (regardless of its morphology) are about one order of magnitude greater than those on silicate and graphite.

Diffusion coefficients k vs $T_s$ can be displayed in several ways according to the law used to describe them, namely, k may have a) an empirical law built for fitting the experimental values, b) an Arrhenius-law form, with an activation energy $E_{diff}$ free to vary, or c) a quantum-tunnelling form with a width and height of the barrier. A detailed analytical or numerical solution of the dependence of k with $T_s$ can help have some insight into the physical nature of the diffusive process at play.

Case a), the empirical law we used for fitting the diffusion coefficients as a function of surface temperature, has the form

$$k_{emp}(T) = k_0 + \alpha(T/10)^\beta. \tag{1}$$

The diffusion coefficients given by eqn (1) and eqn (2) provide the diffusion probability of exploring a fraction of 1 ML per unit time, and can be converted into the usual units $cm^2\ s^{-1}$ (Fig. 3, Fig. 4, and Fig. 6) by assuming that there are $10^{15}$ sites $cm^{-2}$.

Fig. 4 displays a fit of the experimental values obtained on amorphous silicate according to the empirical law given in Eq. (1). $k_0$ ($s^{-1}$) can be considered the minimum value of k, or the value k must tend to near T = 0 K. α is a free parameter with values between 0 ad 1 $K^{-1}\ s^{-1}$, and it accounts for diffusion efficiency differences between the various substrates. α is one for water ice while is about 0.1 for graphite and silicate. The dependency on the surface temperature is governed by the factor $(T/10)^\beta$, the exponent β can have a value between 3 and 4, with variations due to the surface nature, although the best fits give typical values of β ~ 3.5.

Case b), the classical Arrhenius law used to model the diffusion coefficients k is

$$k_{Arr}(T) = \nu_0 \exp[-E_{diff}(T)/T]. \qquad (2)$$

$E_{diff}$ is the diffusion barrier expressed in kelvins (eV/$k_b$) and $\nu_0$ (= $10^{12}$ s$^{-1}$), the pre-exponential factor, can be seen as a trial frequency for attempting a new event. In Fig. 5 we present a fit of the diffusion coefficients k on non-porous ASW obtained by using the Arrhenius law. Fig. 5 actually displays the activation energies for diffusion ($E_{diff}$) as a function of temperature. In fact, according to Eq. (2), a suited set of $E_{diff}$ can be used to derive one diffusion coefficient for each temperature. It is thus possible to link each of these diffusion coefficients to an Arrhenius behaviour, and find one energy barrier at each temperature as shown in Fig. 5 (see also dashed lines in Fig. 3). It should be noted, however, that an Arrhenius-law form in which $E_{diff}$ is fixed (indipendent of T), or where a distribution of $E_{diff}$ is given, is not able to fit the data. This is why we discarded the Arrhenius-type behaviour of k as it made no physical sense to us. In fact, a systematic increase of the Arrhenius barrier with temperature seemed to us an *ad hoc* solution. Also, this implies that at low temperatures (~ 6 K) diffusion occurs through low diffusion barriers (e.g., $E_{diff}$ = 170 K). If such low barriers actually exist, they represent fast connections between adsorption sites. Why then would these low energy barriers vanish at higher temperatures? To put it in other terms, why and how atoms would diffuse through slow pathways (high diffusion barriers) at high temperatures (~ 20 K), if faster pathways exist? We consider this unlikely and not physically reasonable.

In Fig. 6 we show a comparison between the classical behaviour (described by an Arrhenius-type law) of the diffusion coefficients as reported by *Karssmeijer et al*[47] for CO molecules on hexagonal water ice, and the trend that we find experimentally for O atoms on amorphous silicate and crystalline water ice. It is clear that our experimental values do not follow an Arrhenius behaviour, suggesting that a classical description is incomplete to explain the experimental data (squares and triangles in Fig. 6). In fact, in a pure thermal diffusion the slope is very different, and if we fit the data by using a classical Arrhenius law, we find values of $\nu_0$ and $E_{diff}$ not physically acceptable. Therefore, a quantum mechanical approach ought to be used to account for the deviations from the classical trend. Our results on oxygen atoms are consistent with a tunnelling-dominated diffusion found for H atoms on $H_2O_{(np)}$ by *Senevirathne et al*[48] in the 6 – 13 K temperature range (the slopes of H- and O-diffusion constant behaviours are similar). They also found that diffusion of H atoms is enhanced around 13 K, as occurs to O atoms around 22 K, just where classical thermal motion begins to predominate over quantum processes.[39,48] *Hama et al*[49] found that this temperature border between quantum and classical diffusion of H atoms is likely to be at $T_s <$ 10 K. We found that at very low temperatures the diffusion of O atoms is better simulated by quantum tunneling through a square barrier.[39,50] The physical parameters we use to describe such a quantum jump are the width *a* and the height $E_a$ of the barrier. The choice of a square barrier, the simplest shape of a potential, was made on purpose to show that the right trend is obtained if one uses quantum-tunneling diffusion, not because we believed that a square barrier was the right one. We believe that any other more realistic potential shape we could use, would not fundamentally change the results, and it would still be unrealistic given the complexity of the distribution of diffusion barriers. We did not try to obtain the best fit of our data, but tried to show that the right trend is obtained if one uses quantum-tunneling diffusion (see solid lines in Fig. 3). Hence, we chose to model the quantum diffusion with two parameters which have a simple physical meaning, although they correspond to macroscopic values that come from the interplay of many microscopic different situations.

The values of k and of all the parameters used to fit the diffusion coefficients on each substrate, using the three methods, are listed in Table 1.

The diffusion coefficients of O atoms calculated on water ices are one order of magnitude greater than those found on amorphous silicate and oxidised HOPG, namely O diffusive mechanism is more efficient on icy grains. Also, as opposed to the case of H atoms, there is no difference between the efficiency of O mobility on the three types of water ices investigated ($H_2O_{(p)}$, $H_2O_{(np)}$, and $H_2O_{(c)}$). In the light of our experimental results, we can only observe and simply report this finding. In fact, to deal with atoms makes it very difficult to derive key parameters such as the energy barrier for diffusion, or even the energy barrier for desorption, hence no pertinent assumption can be made to explain these findings from a physicochemical point of view. However, to give a physical explanation of our results is beyond the scope of this paper, since we believe that quantum calculations and simulations will be necessary to thoroughly describe O diffusion mechanisms at low temperatures.

**Astrophysical implications**

As far as the diffusion of O atoms is concerned, it turns out that, whenever a diffusive process exists, this has an impact on the chemistry occurring at the surface of dust grains. In fact, either the formation of some species may be enhanced or at least the relative abundances of the final products is affected if O diffusion is efficient. An important example of how O-atom mobility can module the abundances of key species of ices in the ISM is the case of the $H_2O/CO_2$ ratio.

In dense quiescent molecular clouds, hydrogen atoms have always been thought to be the only mobile species on the surface of icy grains. Most of the molecular variety observed in interstellar ices has long been considered the outcome of H-atom addition reactions involving O, $O_2$, $O_3$, CO, N, and NO. Water formation, for example, is the final and most stable species of all the chemical network between H and O, $O_2$ and $O_3$, which justifies its role as the most abundant ice in the Universe. If the reactive partner of H is CO, then $CH_3OH$ is obtained via a series of successive hydrogenations. On the other hand, if H is the only mobile species able to scan the entire surface of the grain,[21] it may be difficult to explain the abundance of $CO_2$, the second most abundant condensed species. $CO_2$ can also be formed via energetic processes by irradiating ice mixtures of $H_2O$ and CO with UV photons or ions. In the dense core of molecular clouds, however, these processes may not apply, and $CO_2$ can only be formed via non-energetic mechanisms, i.e., the reactions CO+OH and CO+O. If these chemical routes leading to $CO_2$ involved only species not mobile at 10 K, then $CO_2$ formation would be greatly hindered by the rate of accretion and the high mobility of H atoms, able to reach CO, OH, and O long before these species can meet to form carbon dioxyde. Our present and previous works introduce strong arguments to believe that O atoms too are mobile at very low temperatures. This implies that the formation rate of $CO_2$ in dense clouds is governed by a balance between the accretion rate of H atoms and the diffusion rate of O atoms on the surface of dust grains. The cartoon in Fig. 7 shows that when the accretion rate of H atoms is dominant, $H_2O$ and $CH_3OH$ are for the most part the final products; when the diffusion rate of O atoms prevails, formation of $CO_2$ (and $O_3$) is favoured.

With this in mind, we made some calculations to show the evolution of the relative abundances of H atoms and O atoms on the surface of dust grains and – assuming that both species are mobile at low temperaures – how this balance can affect the chemistry within interstellar clouds of various densities. In fact, different environments are characterized by different densities, the abundances of species in the gas phase change and this entails a change in the accretion time-scales of particles on dust grains. In diffuse clouds, hydrogen is mainly in its atomic form and is by far the most abundant atomic species. In dark clouds, hydrogen is mainly in its molecular form, so H atoms become a rather rare reactant with $[H]/[H_2] \sim 10^{-3}$ (see, e.g., *Li & Goldsmith*).[51] The number density of H atoms is mostly governed by the destruction of $H_2$ due to cosmic rays. This value, almost indipendently of the density of the cloud, is of the order of 1 H $cm^{-3}$. On the other hand, the $[O]/[H_2]$ ratio remains approximately constant ($10^{-4}$), thus the number of atomic O, unlike H, is proportional to the density of the cloud (see, for example, Table 1 of *Caselli et al*).[52] For a cloud with number density of $10^4$ cm$^{-3}$, the [H]/[O] ratio is ~ 1/0.75, while for a denser cloud with a density of $10^5$ cm$^{-3}$, the [H]/[O] ratio is ~ 1/7. Therefore, for very dense clouds, O is the most abundant species in atomic form, can accrete on grains and, provided O atoms are mobile, subsequently react with other species before these get fully saturated by H-additions. Accretion rates of H atoms and diffusion coefficients of O are then the key factors to be compared in order to determine at what density of the medium oxidation reactions become comparable to H-atom additions.

In Fig. 8 we show the time interval between two impacts of particles of the same species (H or O) on a single dust grain, as a function of the density $n$ of the cloud. The time intervals between two arrivals are derived from the actual particle flux of a given species. The interstellar flux of species accreting on dust grains can be calculated as follows:

$$\Phi_x = 1/4 \; n_x \; v_x \qquad (3)$$

where $n_x$ is the density of species x in gas phase and $v_x = (8k_bT/\pi m_x)^{0.5}$ is its mean velocity. $\Phi_x$ is thus expressed in particles cm$^{-2}$ s$^{-1}$. For our calculation, we can approximate the dust grains to spheres with typical radius r = 0.1 μm, with accessible surface area $A = 4\pi r^2$. The time interval between the impacts of two particles then is

$$t = (n_x \; v_x \; A/4)^{-1}. \qquad (4)$$

In Fig. 8, the grey solid line rappresents the time interval between the impact of two hydrogen atoms, calculated by assuming a constant density of H atoms $n_H$ = 2.3 cm$^{-3}$ (from *Li & Goldsmith*).[51] The density of O atoms is proportional to the density of the clouds $n$, namely, $n_O = 5 \cdot 10^{-4} \, n$. The time interval between the arrival of two O atoms is displayed as a red solid line, which clearly shows that arrivals of oxygen atoms become more frequent (shorter time between two impacts) with the density of the cloud. The grey and red lines cross around a density $n$ of $10^4$ cm$^{-3}$. This suggests that for cloud densities of $\sim 10^4$ cm$^{-3}$ the accretion rates of H and O are comparable and, given that both species can diffuse, oxidation reactions on grains may play a role, although H-atom additions are still dominant owing to the higher mobility of H. In Fig. 8 we also indicate the mean time O atoms need for completing a scan of all the adorption sites on the surface of one typical grain used above, with radius = 0.1 μm and $10^6$ absorption sites ($10^{15}$ sites cm$^{-2}$). Mean times needed for a complete scan of the grain surface were calculated for a surface temperature of 10 K by using the diffusion constants k of O atoms on each substrate presented in this work, taking into account that k = $10^{-15}$ cm$^2$ s$^{-1}$ corresponds to one jump per unit time. As to H atoms, the mean time for scanning the entire surface of water ice was derived by the energy barrier for diffusion of 255 K (at 10 K) given by *Matar et al.*[21] Again, in Fig. 8, it is interesting to observe the intersection occurring at $n \sim 10^5$ cm$^{-3}$ between the red line and the band giving the mean time H atoms employ to scan the whole surface of the grain. This implies that at cloud densities of $\sim 10^5$ cm$^{-3}$ or greater the diffusion and accretion rate of H atoms are smaller than the accretion rate of O atoms. Therefore, in very dense clouds oxygen atoms may become the dominant reaction partner able to react with CO and produce $CO_2$, as well as with H and produce OH. Since H atoms are rare in this environments, OH will not be readily transformed into water via hydrogenation, and also the hydroxil radical is likely to react with the abundant CO molecules to form $CO_2$.

**Acknowledgements**
The LERMA-LAMAp team in Cergy acknowledges the support of the national PCMI programme founded by CNRS, and the Conseil Regional d'Ile de France through SESAME programmes (contract I-07597R). MM acknowledges financial support by LASSIE, a European FP7 ITN Communitys Seventh Framework Programme under Grant Agreement No. 238258. MM also thanks Prof. Tomellini (Università di Roma Tor Vergata) for fruitful discussions.

**Table 1** Best fit parameters of the three methods used to model the diffusion constants for O diffusion on five different grain surface analogues.

|  | Quantum tunnelling | | Arrhenius law | Empirical law | | |
| --- | --- | --- | --- | --- | --- | --- |
|  | $a$ (Å) | $E_a$ (K) | $E_{diff\,[6<T_s<25]}$ (K) | $k_0/10^{-15}$ | $\alpha$ | $\beta$ |
| Porous ASW | 0.69±0.10 | 530±70 | 170 < $E_{diff}$ < 600 | 1.30 | 1 | 3 |
| Non-porous ASW | 0.70±0.05 | 520±60 | 170 < $E_{diff}$ < 600 | 1.21 | 1 | 3 |
| Crystalline water ice | 0.69±0.05 | 500±50 | 170 < $E_{diff}$ < 600 | 1.42 | 1 | 3 |
| Amorphous silicate | 0.67±0.10 | 720±70 | 290 < $E_{diff}$ < 740 | 0.15 | 0.1 | 4 |
| Oxidised HOPG | 0.67±0.10 | 740±60 | 290 < $E_{diff}$ < 740 | 0.1 | 0.1 | 4 |

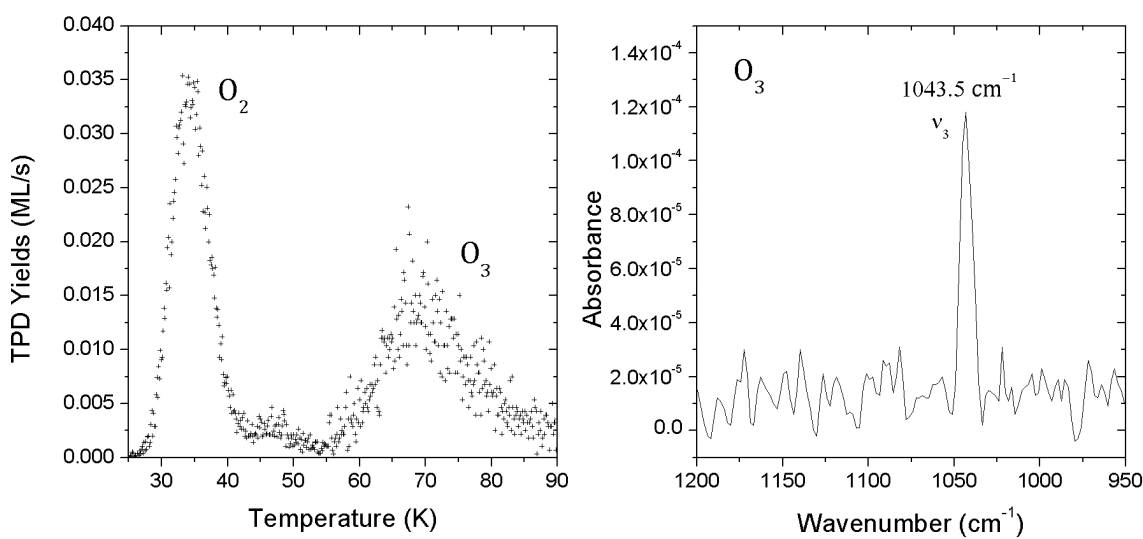

**Fig. 1**  Left panel: $O_2$ and $O_3$ TPD traces obtained after deposition of 0.3 ML of $O+O_2$ on oxidised HOPG held at 6.5 K. Right panel: RAIR spectrum recorded after deposition of 0.3 ML of $O+O_2$ on oxidised HOPG at 6.5 K; the absorption band at 1043.5 cm$^{-1}$ is due to the $\nu_3$ asymmetric stretching mode of $O_3$.

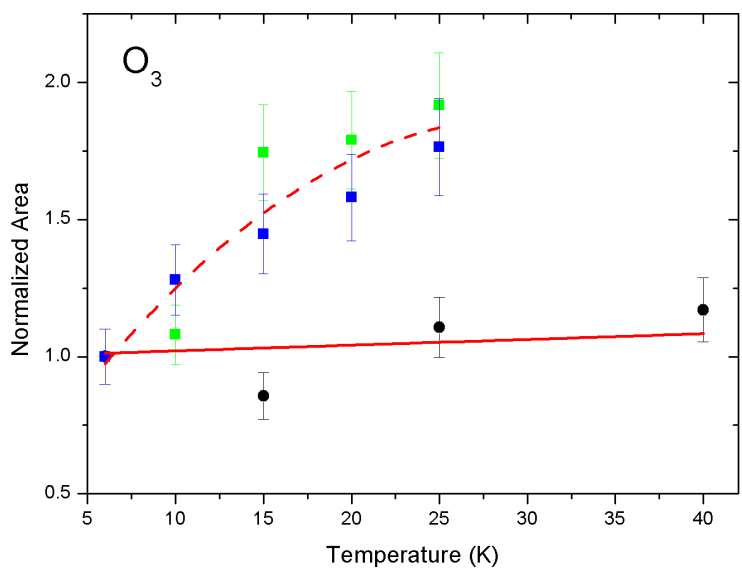

**Fig. 2** Ozone yields derived from TPD peak areas (blue squares) and RAIRS ozone $\nu_3$-band area (green squares) after deposition of 0.3 ML of O+$O_2$ on oxidised HOPG at 6.5, 10, 15, 20, and 25 K. Black circles represent RAIRS ozone $\nu_3$-band areas obtained by depositing a unique 0.3-ML dose of O+$O_2$ on oxidised HOPG held at 6.5 K then by recording a RAIR spectrum at different surface temperatures (6.5, 15, 25, and 40 K). The dashed and the solid red lines are fits of the experimental values and serve as a guide to the eye. All ozone yields were normalised to the $O_3$ yield obtained from oxidised HOPG after deposition of 0.3 ML O+$O_2$ at 6.5 K.

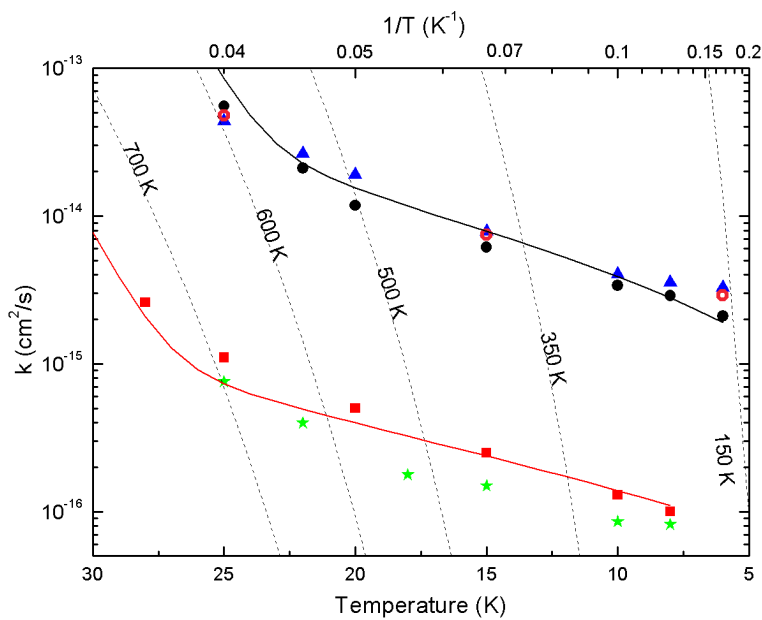

**Fig. 3**     Diffusion constants k of O atoms obtained on $H_2O_{(p)}$ (open pink circles), on $H_2O_{(np)}$ (black circles), $H_2O_{(c)}$ (blue triangles), amorphous silicate (red squares), and oxidised HOPG (green stars), plotted as a function of surface temperature. Dashed lines represent a series of Arrhenius-type laws generated by using five values of $E_{diff}$ (from 150 to 700 K). The two solid lines are best fits of the experimental values obtained through the quantum-tunneling diffusion law for O atoms on $H_2O_{(np)}$ (black solid line) and amorphous silicate (red solid line); see Table 1 for best fit values of $a$ (barrier width) and $E_a$ (barrier height).

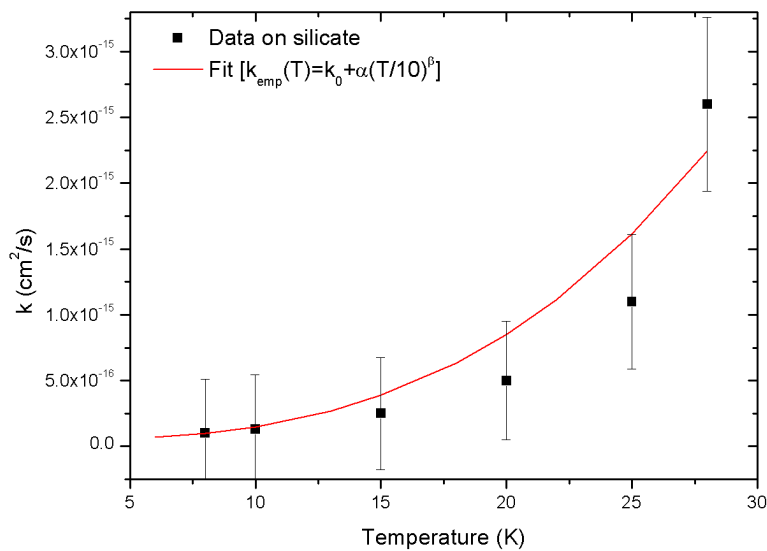

Fig. 4     Black squares represent diffusion constants of O atoms on amorphous silicate as a function of temperature. The red solid line is a best fit of diffusion constants vs temperature obtained by using the empirical law given in Eq. (1); see Table 1 for best fit values of $k_0$, $\alpha$, and $\beta$.

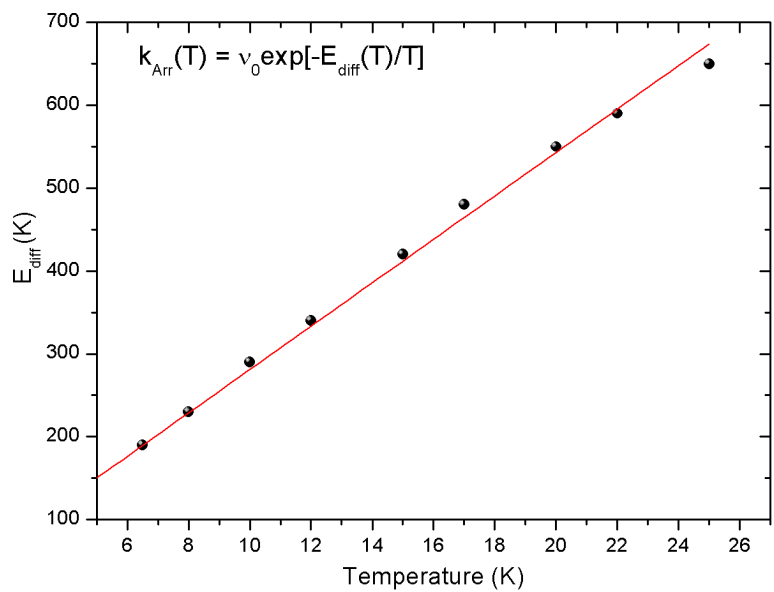

**Fig. 5**   Energy barrier for diffusion on $H_2O_{(np)}$ as a funcion of surface temperature in the case diffusion constants are derived from the Arrhenius-type law given in Eq. (2). The red solid line represents a linear fit of $E_{diff}(T)$. A single value of $E_{diff}$ cannot be a solution satisfying the whole set of diffusion constants observed in the 6 – 25 K temperature range; see Table 1 for the interval of $E_{diff}$ values needed to obtain $k_{Arr}$ between 6 and 25 K.

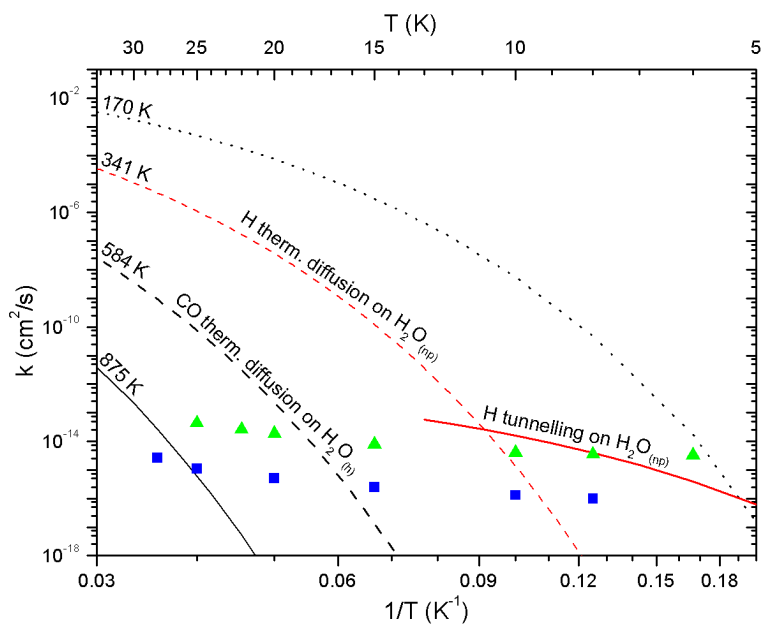

**Fig. 6** Comparison between diffusion behaviours of CO molecules, H atoms, and O atoms. Blue squares and green triangle represent O-atom diffusion constants as a function of surface temperature on amorphous silicate and crystalline water ice, respectively (this work). The black dashed line represent the thermal diffusion (Arrhenius behaviour) of CO molecules on hexagonal water ice found in *Karssemeijer et al.*[47] The red solid line and the red dashed line display the H-atom tunnelling (6 – 13 K) and the H thermal diffusion ($T_s > 13$ K), respectively, obtained by *Senevirathne et al*[48] on compact amorphous water ice. The difference between the slopes of CO and O behaviours, and the similarity between the slopes of H tunnelling and O data, corroborates the conclusion that O atoms diffuse via quantum tunnelling in the surface temperature domain between 6 and 22 K.

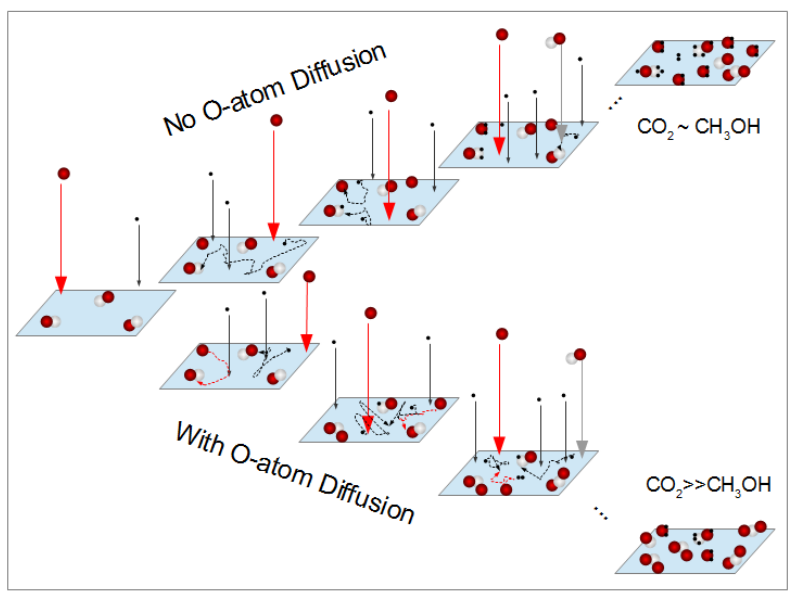

**Fig. 7**   Schematic view of the two possible scenarios concerning the $H_2O/CO_2$ balance on icy grains in dense molecular clouds. If H atoms (black dots) are the only mobile species, H-addition mechanisms are dominant and the formation of water (H + OH/O/O$_2$/O$_3$ → $H_2O$) and of other H-saturated species (e.g., CO + 3H → $CH_3OH$) is favoured. On the other hand, if also O atoms (red circles) can diffuse at very low temperatures, the formation of $CO_2$ in dense clouds may proceed via non-energetic reactions (CO+O and CO+OH) as well, making possible that $CO_2$ is the second most abundant ice in the ISM.

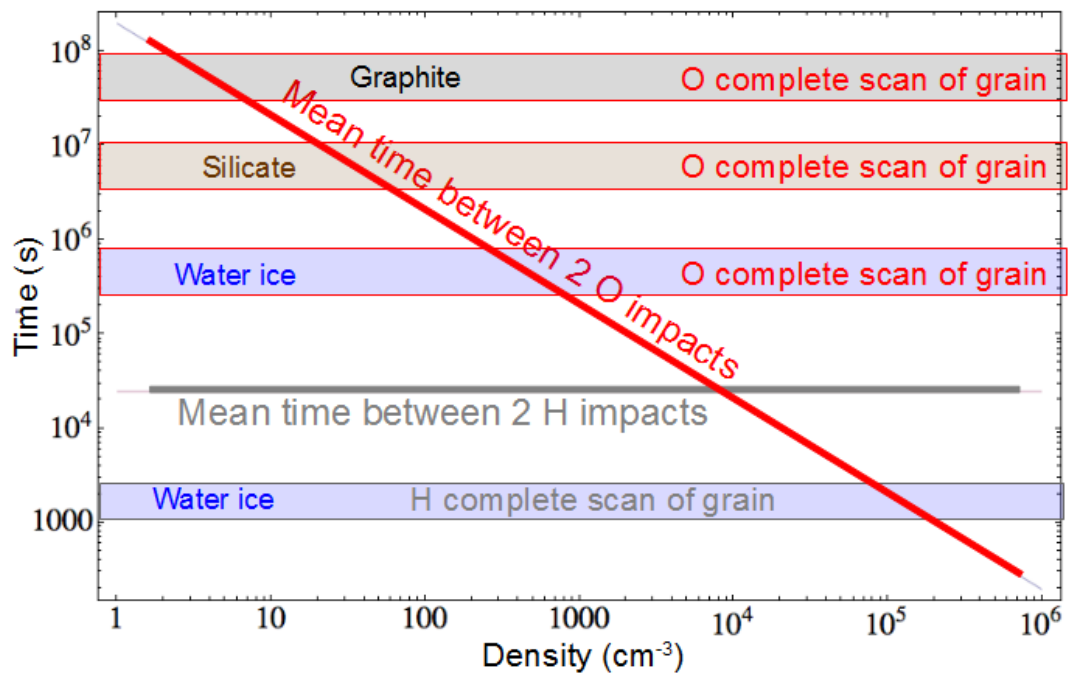

Fig. 8     Time intervals between two impacts of H and O, and times employed to scan a whole grain (shaded horizontal bands) on various surfaces of interest, are plotted as a function of the density of the cloud. The time interval between two arrivals of H is constant as the density of H atoms remains rather constant regardless of the density of the medium, while O atoms abundance grows with cloud density.